\documentclass[twocolumn,10pt,elsearticle]{IEEEtran}

\usepackage{amsmath}
\usepackage{theorem}
\usepackage{amsmath}
\usepackage{mathrsfs}
\usepackage{epsfig}
\usepackage{amsfonts}
\usepackage{amssymb}
\usepackage{mathrsfs}
\usepackage{euscript}
\usepackage{graphicx}
\usepackage{multirow}
\usepackage{cite}

\begin{document}

\title{Cost-Bandwidth Tradeoff In Distributed Storage Systems}
\author{
\authorblockN{Soroush Akhlaghi, Abbas Kiani and Mohammad Reza Ghanavati}\\
\authorblockA{Department of Electrical Engineering\\
Shahed University \\
Tehran, Iran\\
Email: \{akhlaghi,akiani,ghanavati\}@shahed.ac.ir}
}
\maketitle

\begin{abstract}
Distributed storage systems are mainly justified due to the limited amount of storage capacity and improving the reliability through distributing data over multiple storage nodes. On the other hand, it may happen the data is stored in unreliable nodes, while it is desired the end user to have a reliable access to the stored data. So,~in an event that a node is damaged,~to prevent the system reliability to regress,~it is necessary to regenerate a new node with the same amount of stored data as the damaged node to retain the number of storage nodes, thereby having the previous reliability.

This requires the new node to connect to some of existing nodes and downloads the required information, thereby occupying some bandwidth, called the repair bandwidth. On the other hand, it is more likely the cost of downloading varies across different nodes. This paper aims at investigating the theoretical cost-bandwidth tradeoff, and more importantly, it is demonstrated that any point on this curve can be achieved through the use of the so called generalized regenerating codes which is an enhancement of the regeneration codes introduced by Dimakis et al. in~\cite{dim1}.

\end{abstract}



\section{Introduction}
Data in distributed storage systems should be stored reliably for a long period of time. This is due to the need for surviving in the case that individual failures occur, thus having a long-term durability. To this end, the system should have the possibility of self-repairing in the case that a node is failed or leaves the system. This requires a great deal of data transferring due to repairing a failure node, called repair bandwidth. In some cases, a great deal of repair bandwidth is consumed to construct a new node.

To have a reliable data, various strategies have been proposed which basically attempt to add some redundancy bits to the original data and distributing the encoded data across distinct nodes in an effective manner. The simplest strategy is replication in which each node stores the original data file, hence, the data of one node is adequate to reconstruct the original data. However, this is not a wise method due to the need to a high storage capacity.~To address this issue,~in~\cite{era1,era2} instead of exploiting naive replication code, an erasure coding is used in which the original data file of size $M$ is divided into $k$ pieces of size $M/k$, and encoded into $n$ data fragments to be stored in one of existing $n$ nodes. The encoding process is such that having access to the stored data of $k$ nodes is adequate to reconstruct the original data. In other words, a new node should be connected to $k$ nodes to have an access to all information. As a result, for a large value of $k$, the storage capacity of each node is dramatically reduced as compared to the replication code, since instead of storing data size of $M$, we need to merely store a fragment of data size $M/k$ at each node~\cite{erarep2,erarep1}. Although, the erasure code requires the same repair bandwidth as compared to the replication code and imposes a decoding complexity into the system, it makes a balance between the system reliability and redundancy.

To take the advantages of both replication~(simple decoding method) and erasure coding~(low storage capacity), in~\cite{erarep2} a hybrid strategy is proposed. This strategy uses a single node containing an exact replica of the original data file as well as some nodes with the structure of erasure coding. Thus, for generating a new data fragment, this replica is used and just a data of size $M/k$ is transferred across the network. Although the repair bandwidth of the hybrid strategy is reduced, the system complexity is greatly increased, i.e.,~if the replica is failed, creating a new fragment is deferred until restoring the replica. This in turn, may not be feasible when there is a stringent delay constraint.

This motivated Dimakis et al. in~\cite{dim1} to deduce an elegant coding strategy, dubbed regenerating codes~(RC), to reduce the repair bandwidth without the use of replica. It is shown that for creating a new data fragment, the newcomer node should be connected to $d$ nodes ($d\ge k$) and download $\beta$ bits from each surviving nodes. Accordingly, a trade-off between storage per node and repair bandwidth ($d\beta$) is identified.

Regenerating codes and other existing methods are motivated by the assumption that surviving nodes have equal download cost, and creating a new node is accomplished through downloading the same amount of information from each surviving node. However, it may happen there is a different cost associated with each node. Thus, in an attempt to replace a damaged node with a new node, one may want to make a balance between the download cost and the repair bandwidth.

The current study aims to address the aforementioned issue when there are two sets of nodes, each having different download costs, while the nodes of each set have the same cost. However, the material in this paper can be readily extended to more general cases. Accordingly, it is assumed a newcomer node downloads~$\beta_1$ and $\beta_2$ bits, respectively, from each surviving node of cost $C_1$ and $C_2$, where it is simply assumed $C_1\leq C_2$. It will be later shown that under certain conditions, if $\beta_1$ is larger than $\beta_2$, the total download cost is reduced at the expense of increasing the repair bandwidth. In other words, the more $\beta_1$ is larger than $\beta_2$, the less download cost is produced, while having more repair bandwidth as if $\beta_1=\beta_2$. Moreover, for a given $\beta_1$ and $\beta_2$, congruent to what is done in~\cite{dim1}, a trade-off between the storage per node and repair-bandwidth is identified.

The rest of paper is organized as follows: In section~\ref{sec:backgroung}, distributed storage systems are briefly introduced and their equivalent \emph{Information Flow Graph} is introduced. Accordingly, it is argued that network coding can approach the capacity of such systems. Finally, regenerating codes are motivated and briefly introduced. Section~\ref{sec:model} states the problem formulation and motivates the main idea, finally gives an overview of the approach. Sections~\ref{sec:numerical},\ref{sec:conclusion}, present numerical results and conclude the paper, respectively.

\section{Background}\label{sec:backgroung}
\subsection{Distributed Storage Systems and connection to Network Coding}
\begin{figure}
\centering
\epsfig{file=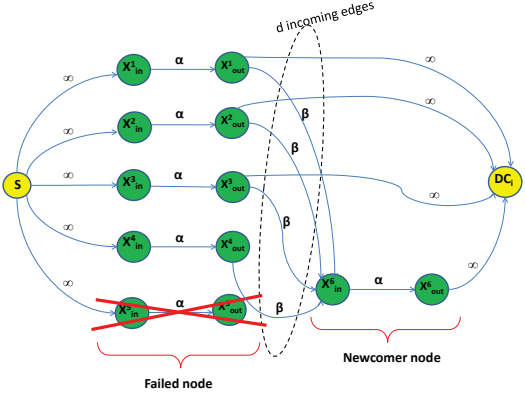,width=1\linewidth,clip=}
\caption{An example of \emph{Information Flow Graph} when a failure is occurred (it is marked by cross lines), thus a new node is initiated.}
\label{fig:0}
\end{figure}
In distributed storage systems, nodes join or leave the network continuously, hence, the network configuration varies across time. Motivated by the pioneering work in~\cite{dim1}, this network can be thought as an~\emph{information flow graph}, a directed acyclic graph consisting of three types of nodes: (i) A single source node ($S$), (ii) Some intermediate nodes and (iii) Data collectors ($DC$ nodes). The source node is the source of original data file, intermediate nodes are storage nodes and each data collector corresponds to a request for reconstructing original data file. Each storage node is represented by pairs of incoming and outgoing nodes connected by a directional edge whose capacity is the corresponding storage capacity of this storage node. In this work, we simply assume all storage nodes are of capacity $\alpha$. Moreover, it is assumed edges departing the storage nodes and arriving to a DC node have infinite capacity. This reflects the fact that DC nodes have access to all stored data of the surviving nodes they are connected to.

As is mentioned earlier, the corresponding~\emph{information flow graph} evolves constantly across time to reflect any changes happening throughout the network. This graph starts from the source node,~indicating it is the only active node at the first step. Then, assuming the total number of storage nodes is $n$,~the source node divides the original data file of size $M$ into $k$ pieces, encodes these $k$ pieces to $n$ data fragments each to be stored in one of existing storage nodes through direct edges of infinite capacity. In the case that a storage node leaves the system or a failure occurs, this node is replaced by a new one, called the newcomer node. The newcomer connects to $d$ active nodes out of $n-1$ existing nodes and downloads~$\beta$ bits from each. Accordingly, the corresponding~\emph{information flow graph} is updated through establishing $d$ directed edges of capacity $\beta$, starting from outgoing nodes affiliated to the selected storage nodes and terminating to the corresponding incoming node of the newcomer (Fig.\ref{fig:0}). In this case, the total information received by the newcomer node,~$d\beta$, is called the repair bandwidth~$(\gamma)$.
Finally, the data is reconstructed at each DC node through connecting to any arbitrary set of $k$ nodes (storage nodes), including the newcommer nodes. The edges connecting the selected storage nodes to the corresponding DC node are assumed to be of infinite capacity.

Incorporating the graphical representation of distributed storage systems gives the opportunity to relate the storage capacity as well as repair bandwidth of the original problem to some characteristics of the corresponding~\emph{information flow graph}. Specifically, we are interested in an important quantity, call the network throughput introduced by Ahlswede et al. in~\cite{swed}, which basically identifies the maximum allowable information flow from a source to a destination node, assuming each link is subject to a limited capacity. Accordingly, it is demonstrated that using a proper coding at intermediate nodes, it is possible to get the information with a throughput at most equal to what is promised by the so called min-cut theorem~\cite{swed}. This is achieved through using an elegant coding strategy, called network coding, which basically can approach the multicast capacity of such networks~\cite{lnc,lnc2}. The notion of using network coding has beaten the previous belief of using simple routing mechanism at intermediate nodes.

\subsection{Regenerating Codes}
As is mentioned earlier, for erasure coding, having an access to the data of $k$ storage nodes out of existing $n$ nodes is adequate to reconstruct the original data file. Thus, the newcomer needs to connect to exactly $d=k$ nodes and downloads all of stored data ($\alpha=M/k$), thus~$\beta=\alpha=M/k$. So the repair bandwidth becomes the same as the size of data file, i.e.,~$\gamma=d\beta=M$. On the other hand, Dimakis et al. in~\cite{dim1} show that if a newcomer could connect to more than $k$ surviving nodes and downloads a certain function of their stored information, a lower repair bandwidth would be achieved, while having the same storage capacity as compared to that of erasure coding. 

To this end, it is shown the task of computing the repair bandwidth can be translated to a multicast problem over the corresponding information flow graph for which an optimal trade-off between the storage per node,~$\alpha$, and the repair bandwidth,~$\gamma$, is identified. This optimal trade-off curve includes two extremal points corresponding to the minimum storage capacity per node and minimum repair bandwidth, respectively. Recall that any points on the trade-off curve, including the extremal points can be achieved by the use of network coding approach. The former, minimum storage capacity, is achieved by use of the so called Minimum Storage Regenerating (MSR) codes. The latter, is realized through using Minimum Bandwidth Regenerating (MBR) codes. Accordingly, the corresponding storage capacity per node ($\alpha$) and repair bandwidth ($\gamma$) for MSR and MBR codes are computed as follows~\cite{dim1}:
\begin{eqnarray}\label{equ1}
(\alpha_{MSR},\gamma_{MSR})\!\!\!&=&\!\!(\frac{M}{k},\frac{Md}{k(d-k+1)})\nonumber\\
(\alpha_{MBR},\gamma_{MBR})\!\!\!&=&\!\!(\frac{2Md}{2kd-k^{2}+k},\frac{2Md}{2kd-k^{2}+k})~,
\end{eqnarray}
where in~(\ref{equ1}), it is assumed the total data file is of size $M$. Moreover, $d$ denotes the number of storage nodes which a newcomer is connected to ($d\geq k$), and $k$ represents the total number of nodes which are required to reconstruct the original data file. In other words, a DC node needs to connect to exactly $k$ storage nodes to reconstruct the original data file.

\section{Problem formulation and the proposed method}\label{sec:model}
MSR and MBR codes are motivated by the assumption that the download cost of all storage nodes are the same. However, we rely on a more realistic situation in which storage nodes are subject to different download costs and the download cost is of great concern. Specifically, we concentrate on the case that there are totally two sets of storage nodes $S_1$ and $S_2$ with download costs per information bit equal to $C_1$ and $C_2$, respectively\footnote{This enables the problem can be mathematically tractable. However, one can readily follow the same approach for more general cases.}. Accordingly, in regenerating codes, a newcomer connects to $d$ nodes, each belongs either to $S_1$ or $S_2$. Assuming $d_1$ nodes are of cost $C_1$ and $d_2=d-d_1$ nodes are of cost $C_2$, thus the total cost for reconstructing a damaged node becomes:
\begin{equation}\label{equ3}
C_T=(C_1d_1+C_2d_2)\beta~,
\end{equation}
where $\beta$ is the total information downloaded from each node. Equation~(\ref{equ3}) indicates that the same amount of information is downloaded from each node, no matter which set it basically belongs to. However, an important enquiry may arise; How to make a balance between the repair bandwidth and the total cost?. In this work, we aim at addressing the aforementioned issue and more importantly, to establish a trade-off between the repair bandwidth, the storage capacity, and the total cost.

We employ a variation of the regenerating code, dubbed Generalized Regenerating Code (GRC), in which the newcomer downloads different amount of information depending on the type of storage node. In the course of downloading, we consider there are totaly $d_1$ nodes with download cost~$C_1$ and $d_2$ nodes ($d_2=d-d_1$) with download cost $C_2$ ($C_2\geq C_1$), where~$\beta_1$ and $\beta_2$ bits are downloaded from each of these nodes, respectively. Noting~$C_2\geq C_1$, one can get a lower cost if $\beta_1\geq \beta_2$. Throughout the paper, we assume $\beta_1=k'\beta_2$~\footnote{It is worth mentioning that for some practical purposes, $k'$ should take an integer value.}. As a result, the total cost for constructing a new node in this strategy is as follows:
\begin{equation}\label{equ4}
C_T=C_1d_1\beta_1+C_2d_2\beta_2~.
\end{equation}

It should be noted that, as is shown in the next sections, $k'$ is inversely proportional to the relative download cost, meaning the larger $k'$ results in the less relative cost of GRC as compared to that of the regenerating codes. Then, for a given $k'$,~the problem is translated to computing $\beta_2$ (or equivalently $\beta_1$) for which the minimum repair bandwidth or minimum storage capacity per node is obtained. Accordingly, It is shown even more reduction in $C_T$ is possible at the expense of increasing the repair bandwidth.
In the next section, we examine two different scenarios of $d_1\!\geq\!k$ and $d_1\!<\!k$ to explore the problem.

\section{Scenario A:~$d_1~\geq~k$}\label{sec:scenario A}
Consider any given finite information flow graph $\mathcal{G}$, with a finite set of data collectors. In~\cite{dim1}, it is argued that ``If the minimum
of the min-cuts separating the source with each data collector is larger or equal to the data object size M, then there exists a linear network code defined over a sufficiently large finite field $F$ (whose size depends on the graph size) such that all data collectors can recover the data object''.
\begin{figure}
\centering
\epsfig{file=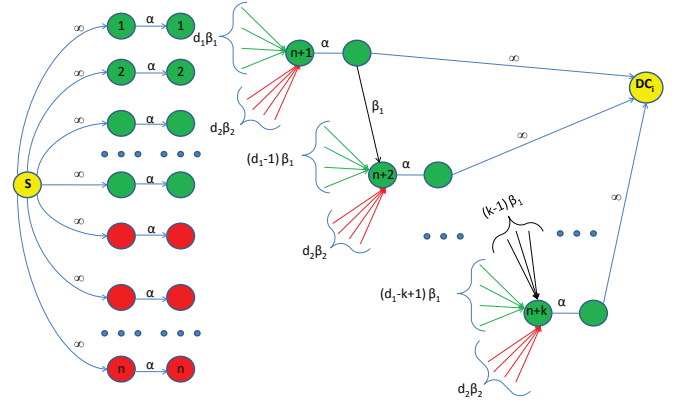,width=1\linewidth,clip=}
\caption{$\mathcal{G}^*$ for $d_1\!\geq\!k$}
\label{fig:1}
\end{figure}
In Fig.\ref{fig:1}, the graph $\mathcal{G}^*$, a portion of the corresponding \emph{Information flow graph~$\mathcal{G}$}, entailing the minimum of the min-cuts for $d_1\!\geq\!k$ is shown. So referring to this flow graph and noting the above argument, the following condition is necessary to reconstruct the original data file:
\begin{equation}\label{equ4}
\sum_{i=0}^{k-1} \textrm{min}\{(d_1\beta_1+d_2\beta_2-i\beta_1),~\alpha\}\!\geq\!M~.
\end{equation}

Thus, using~(\ref{equ4}) and noting $\beta_1=k'\beta_2$, and after some manipulations, a tradeoff between $\alpha_{min}$ (the minimum required storage) and $\beta_2$ is identified as follows,
\begin{eqnarray}\label{equ5}
\alpha_{min}(d_1,d_2,k',\beta_2)\!\!\!&=&\!\!\!
\left\{\begin{array}{ll}
\frac{M}{k}\!\!\!&\textrm{$\beta_2$}\in\big [f(0),\infty\big)\\
\!\!\!\!\!\!&\\\!\!\!\!\!
\frac{2M-g(i)\beta_2}{2(k-i)}\!\!\!&\textrm{$\beta_2$}\in\big [f(i),f(i-1)\big)~,
\end{array} \right.
\end{eqnarray}
where
\begin{eqnarray}
f(i)&\triangleq& \frac{2M}{2k(d_1k'+d_2-kk')+k'(i+1)(2k-i)}\nonumber\\
g(i)&\triangleq& i(2d_1k'+2d_2-2kk'+(i+1)k')~.
\end{eqnarray}
Thus, $\beta_{2_{min}}$ (the minimum required download from each node) can be computed as,
\begin{eqnarray}
\beta_{2_{min}}\!\!\!&=&\!\!f(k-1)\nonumber\\
&=&\frac{2M}{k(2d_1k'+2d_2-kk'+k')}~.
\end{eqnarray}
In other words, for any $\alpha\!\geq\!\alpha_{min}(d_1,d_2,k',\beta_2)$, the points $(n,k,d_1,d_2,\alpha,\beta_1,\beta_2)$ with linear network coding are achievable.

Thus, the tradeoff curve between the storage capacity ($\alpha$) and the repair bandwidth ($\gamma=\beta_1d_1+\beta_2d_2$) can be established through using (\ref{equ5}), where $\beta_1=k'\beta_2$. This curve has two extremal points. One corresponds to minimum storage capacity and the other related to the minimum repair bandwidth. We call the codes that achieve these points as Generalized Minimum Storage Regenerating~(GMSR) and Generalized Minimum Bandwidth Regenerating~(GMBR) codes, respectively. GMSR is identified with the following storage capacity-repair bandwidth pair,
\begin{eqnarray}\label{equ6}
(\alpha_{\textrm{GMSR}},\gamma_{\textrm{GMSR}})=\big(\frac{M}{k},\frac{M(d_2+k'd_1)}{k(d_1k'+d_2-kk'+k')}\big)~.
\end{eqnarray}
Similarly, for GMBR, we arrive at the following,
\begin{eqnarray}\label{equ7}
\lefteqn{\!\!\!\!\!\!\!\!\!\!(\alpha_{\textrm{GMBR}},\gamma_{\textrm{GMBR}})=}\nonumber\\
&~~~~\nonumber\\
&\big(\frac{2M(d_2+k'd_1)}{k(2d_1k'+2d_2-kk'+k')},\frac{2M(d_2+k'd_1)}{k(2d_1k'+2d_2-kk'+k')}\big)~.
\end{eqnarray}

It can be verified that for the special case of $k'=1$ and $d=d_1+d_2$, equations (\ref{equ6}) and (\ref{equ7}) become similar to the resulting storage capacity-repair bandwidth pairs of MSR and MBR codes~\cite{dim1}, respectively.
Also, for the case of $k'\rightarrow\infty$, noting $\beta_2=\beta_1/k'$, one can conclude that $\beta_2=0$, hence, the nodes with lower download cost are merely exploited throughout the course of downloading. Accordingly, $(\frac{M}{k},\frac{Md_1}{k(d_1-k+1)})$ and $ \big(\frac{2Md_1}{k(2d_1-k+1)},\frac{2Md_1}{k(2d_1-k+1)}\big)$ are the corresponding storage capacity-repair bandwidth pairs of the resulting GMSR and GMBR codes. As is expected, referring to~(\ref{equ1}), these pairs are similar to that of MSR and MBR codes with $d=d_1$.

\subsection{Comparison between GMSR and MSR when $d_1\!\geq\!k$}\label{sub:msr}
Referring to~(\ref{equ6}) and the resulting storage capacity-repair bandwidth of MSR as is given in~(\ref{equ1}), GMSR and MSR yield the same storage capacity per node. However, they exhibit different repair bandwidth. To have a basis of comparison for the resulting repair bandwidth of GMSR and MSR, we define the bandwidth ratio $\rho_{\textrm{MSR}}(k')$ as follows,
\begin{equation}\label{equ8}
\rho_{\textrm{MSR}}(k')\triangleq\frac{\gamma_{\textrm{GMSR}}(k')}{\gamma_{\textrm{MSR}}}=\frac{(d_2+k'd_1)(d-k+1)}{d(d_1k'+d_2-kk'+k')}~.
\end{equation}
It can be verified that as long as $d\!\geq\!k$ and $k\geq 1$, the derivation of~(\ref{equ8}) with respect to $k'$ is positive. As these conditions hold here, $\rho_{\textrm{MSR}}(k')$ is an increasing function with respect to $k'$ and more importantly, noting $\rho_{\textrm{MSR}}(1)=1$, thus $\rho_{\textrm{MSR}}(k')$ is greater than one for $k'\geq 1$. Thus, the repair bandwidth of GMSR is greater than that of MSR.
Moreover, we define the download cost ratio $\eta_{\textrm{MSR}}(k')$ to compare the  download cost of GMSR to that of MSR, as follows,
\begin{eqnarray}\label{equ9}
\eta_{\textrm{MSR}}(k')&\triangleq&\frac{C_{T_{\textrm{GMSR}}}(k')}{C_{T_{\textrm{MSR}}}}\nonumber\\
&=&\frac{(C_1d_1k'+C_2d_2)(d-k+1)}{(d_1k'+d_2-kk'+k')(C_1d_1+C_2d_2)}~.\nonumber\\
\end{eqnarray}
Note that $\eta_{\textrm{MSR}}(1)=1$. In order to have $C_{T_{\textrm{GMSR}}}$ lower than $C_{T_{\textrm{MSR}}}$, $\eta_{\textrm{MSR}}(k')$ should be a decreasing function, meaning to have a negative derivation with respect to $k'$. As a result, taking derivation of (\ref{equ9}), one can verify that the following condition should be satisfied,
\begin{equation}\label{equ10}
\frac{C_2}{C_1}\geq\frac{d_1}{d_1-k+1}~.
\end{equation}
It is worth mentioning that if the above condition holds, the minimum value of $\eta_{\textrm{MSR}}$ is achieved as $k'$ tends to infinity, i.e., $\eta_{\textrm{MSR}}(+\infty)=\frac{C_1d_1(d-k+1)}{(d_1-k+1)(C_1d_1+C_2d_2)}$.

\subsection{Comparison between GMBR and MBR when $d_1\!\geq\!k$}\label{sub:mbr}
Equations~(\ref{equ1}) and (\ref{equ7}) indicate that the storage per node is equal to the repair bandwidth for both MBR and GMBR codes. As a result, any findings for the corresponding repair bandwidths of MBR and GMBR codes, can also be considered for storage per node as well. In this regard, we define the repair bandwidth ratio $\rho_{\textrm{MBR}}(k')$ as follows,
\begin{eqnarray}\label{equ12}
\rho_{\textrm{MBR}}(k')\triangleq\frac{\gamma_{\textrm{GMBR}}(k')}{\gamma_{\textrm{MBR}}}=\frac{(d_2+k'd_1)(2d-k+1)}{d(2d_1k'+2d_2-kk'+k')}~.
\end{eqnarray}
Obviously, we have $\rho_{\textrm{MBR}}(1)=1$. Again, following the same approach as is done in \ref{sub:msr}, one can readily verify that if the conditions $k\!\geq\!1$ and $k'\!\geq\!1$ hold, $\rho_{\textrm{MBR}}(k')$ is always greater than one. Thus, the repair bandwidth of GMBR is greater than that of MBR. Accordingly, we define the download cost ratio as follows,
\begin{eqnarray}\label{equ13}
\eta_{\textrm{MBR}}(k')&\triangleq&\frac{C_{T_{\textrm{GMBR}}}(k')}{C_{T_{\textrm{MBR}}}}\nonumber\\
&=&\frac{(C_1d_1k'+C_2d_2)(2d-k+1)}{(2d_1k'+2d_2-kk'+k')(C_1d_1+C_2d_2)}~.\nonumber\\
\end{eqnarray}
Note that $\eta_{\textrm{MBR}}(1)=1$. Taking derivation of $\eta_{\textrm{MBR}}(k')$ with respect to $k'$, one can verify that to have $\eta_{\textrm{MBR}}(1)\leq 1$, the following condition should be satisfied,
\begin{equation}\label{equ14}
\frac{C_2}{C_1}\geq\frac{2d_1}{2d_1-k+1}~.
\end{equation}
It is worth mentioning that the minimum value of $\eta_{\textrm{MBR}}$ is achieved as $k'$ tends to infinity, i.e., $\eta_{\textrm{MBR}}(+\infty)=\frac{C_1d_1(2d-k+1)}{(2d_1-k+1)(C_1d_1+C_2d_2)}$.

\section{Scenario B:~$d_1\!<\!k$}
\begin{figure}
\centering
\epsfig{file=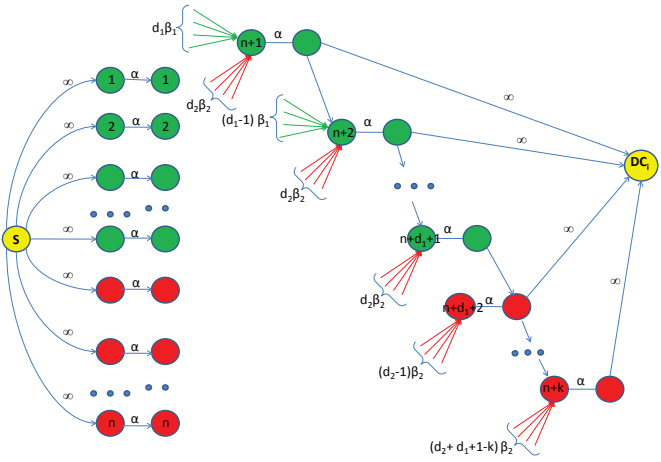,width=1\linewidth,clip=}
\caption{$\mathcal{G}^*$ for $d_1<k$}
\label{fig:4}
\end{figure}

In this case, the information flow graph $\mathcal{G}^*$ has a minimum min-cut similar to what is shown in Fig.\ref{fig:4}.
As a result, according to min-cut theorem as is addressed in Section~\ref{sec:scenario A}, the following condition should be satisfied,
\begin{eqnarray}\label{equ16}
\lefteqn{\sum_{i=0}^{d_1}\textrm{min}\{(d_1\beta_1
+d_2\beta_2-i\beta_1),~\alpha\}+}\nonumber\\
&\nonumber\\
&\sum_{i=d_1+1}^{k-1}\textrm{min}\{(d_1+d_2-i)\beta_2,~\alpha\}~\geq~M
\end{eqnarray}

The above condition introduces a tradeoff between $\alpha$ and $\beta_2$ which is computed as follows,
\begin{eqnarray}\label{equ17}
\alpha_{\textrm{min}}(d_1,d_2,k',\beta_2) =~~~~~~~~~~~~~~~~~~~~~~~~~~~~~~~~~~~~~~\nonumber\\
&\nonumber\\
\left\{\begin{array}{ll}
\frac{M}{k}\!\!&\textrm{$\beta_2$}\in\big[f_1(0),\infty\big)\\
&\nonumber\\
\frac{2M-g_1(i)\beta_2}{2(k-i)}\!\!&\textrm{$\beta_2$}\in\big[f_1(i),f_1(i-1)\big)\\
&\nonumber\\
\frac{2M-(g_1(k-d_1-1)+g_2(i))\beta_2}{2(d_1-i)}\!\!&\textrm{$\beta_2$}\in\big[f_2(i),f_2(i-1)\big)~,\\
\end{array} \right.
\end{eqnarray}
where
\begin{eqnarray}\label{equ18}
f_1(i)\!\!&\triangleq&\!\! \frac{2M}{2k(d-k)+(i+1)(2k-i))}\nonumber\\
f_2(i)\!\!&\triangleq&\!\! \frac{2M}{(2kd-k^2-d_1^2-d_1+k+2d_1k')+ik'(2d_1-i-1)}\nonumber\\
g_1(i)\!\!&\triangleq&\!\! i(2d-2k+i+1)\nonumber\\
g_2(i)\!\!&\triangleq&\!\! (i+1)(2d_2+ik')~.
\end{eqnarray}
Thus, $\beta_{2{\textrm{min}}}$ can be computed as,
\begin{equation}\label{equ22}
\beta_{2_{min}}=f_2(d_1-1)=\frac{2M}{2kd-k^2+k+(d_1^2+d_1)(k'-1)}~.
\end{equation}
Accordingly, GMSR and GMBR, two extremal points of trade-off curve, have the following storage capacity-repair bandwidth,
\begin{eqnarray}\label{equ23}
\!\!\!\!\!\!\!\!\!\!\!\!\!\!\!\!\!\!\!\!\!\!\!\!\!\!\!\!\!\!\!(\alpha_{\textrm{GMSR}},\gamma_{\textrm{GMSR}})~&=&(\frac{M}{k},\frac{M(d_1k'+d_2)}{k(d-k+1)})~.
\end{eqnarray}
\begin{eqnarray}\label{equ24}
\lefteqn{(\alpha_{\textrm{GMBR}},\gamma_{\textrm{GMBR}})~~=~}\nonumber\\
&\nonumber\\
&(\frac{2M(d_1k'+d_2)}{2kd-k^2+k+(d_1^2+d_1)(k'-1)},\frac{2M(d_1k'+d_2)}{2kd-k^2+k+(d_1^2+d_1)(k'-1)})~.
\end{eqnarray}

\subsection{Comparison between GMSR and MSR when $d_1<k$}
Referring to~(\ref{equ1}) and (\ref{equ23}), GMSR and MSR have an equal storage capacity per node. To get an insight regarding the repair bandwidth, we define the following repair bandwidth ratio,
\begin{equation}\label{equ25}
\rho_{\textrm{MSR}}(k')\triangleq\frac{\gamma_{\textrm{GMSR}}(k')}{\gamma_{\textrm{MSR}}}=\frac{d_1k'+d_2}{d}~.
\end{equation}
This ratio is always greater than one for $k'>1$, meaning GMSR code imposes a large bandwidth to the system as compared to MSR. Similarly, the download cost ratio is defined as,
\begin{equation}\label{equ26}
\eta_{\textrm{MSR}}(k')\triangleq\frac{C_{T_{\textrm{GMSR}}}(k')}{C_{T_{\textrm{MSR}}}}=\frac{C_1d_1k'+C_2d_2}{C_1d_1+C_2d_2}~.
\end{equation}
Again $\eta(k')$ for all positive values of $k'$ is greater than one. Having larger repair bandwidth and storage capacity as well as higher download cost, one can conclude that GMSR does not have favorable result as compared to MSR. Thus, GMSR does not perform well for the case of $d_1<k$, meaning in this case it is better to set $\beta_1=\beta_2$ (MSR approach).

\subsection{Comparison between GMBR and MBR when $d_1<k$}
As the storage per node is equal to the repair bandwidth for both MBR and GMBR codes, we concentrate on the repair bandwidth. Again, we define the repair bandwidth ratio $\rho_{\textrm{MBR}}(k')$ as follows:
\begin{eqnarray}\label{equ27}
\rho_{\textrm{MBR}}(k')&\triangleq&\frac{\gamma_{\textrm{GMBR}}(k')}{\gamma_{\textrm{MBR}}}\nonumber\\
&=&\frac{(d_1k'+d_2)(2kd-k^2+k)}{\big(2kd-k^2+k+(d_1^2+d_1)(k'-1)\big)d}~.\nonumber\\
\end{eqnarray}
$\rho(k')$ has a positive derivative with respect to $k'$ and noting $\rho(1)=1$ it follows $\rho(k')\geq 1$ for $k'\geq 1$. Thus, MBR outperforms GMBR in terms of having lower repair bandwidth.~Similarly, we define the download cost ratio as follows,
\begin{eqnarray}\label{equ28}
\eta_{\textrm{MBR}}(k')&\triangleq&\frac{C_{T_{\textrm{GMBR}}}(k')}{C_{T_{\textrm{MBR}}}}\nonumber\\
&=&\!\!\!\!\!\frac{(C_1d_1k'+C_2d_2)(2kd-k^2+k)}{(C_1d_1+C_2d_2)\big(2kd-k^2+k+(d_1^2+d_1)(k'-1)\big)}~.\nonumber\\\!\!\!\!\!\!\!\!\!\!\!\!
\end{eqnarray}
Again, to have the download cost of GMBR lower than that of MBR, the following condition should be satisfied,
\begin{equation}\label{equ29}
\frac{C_2}{C_1}\geq\frac{2kd-k^2+k-d_1^2-d_1}{d_2(d_1+1)}~.
\end{equation}
In this case, the minimum value of $\eta$ is achieved as $k'$ tends to infinity, i.e.,~$\eta(+\infty)=\frac{(C_1d_1)(2kd-k^2+k)}{(C_1d_1+C_2d_2)(d_1^2+d_1)}$
\section{Numerical Results}\label{sec:numerical}
\begin{figure}
\centering
\epsfig{file=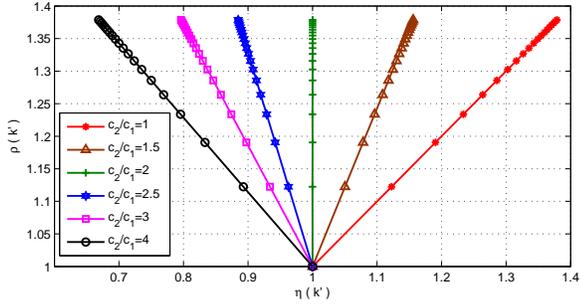,width=1\linewidth,clip=}
\caption{The tradeoff curves between the relative cost and repair bandwidth ratio for GMSR code.}
\label{fig:2}
\end{figure}

\begin{figure}
\centering
\epsfig{file=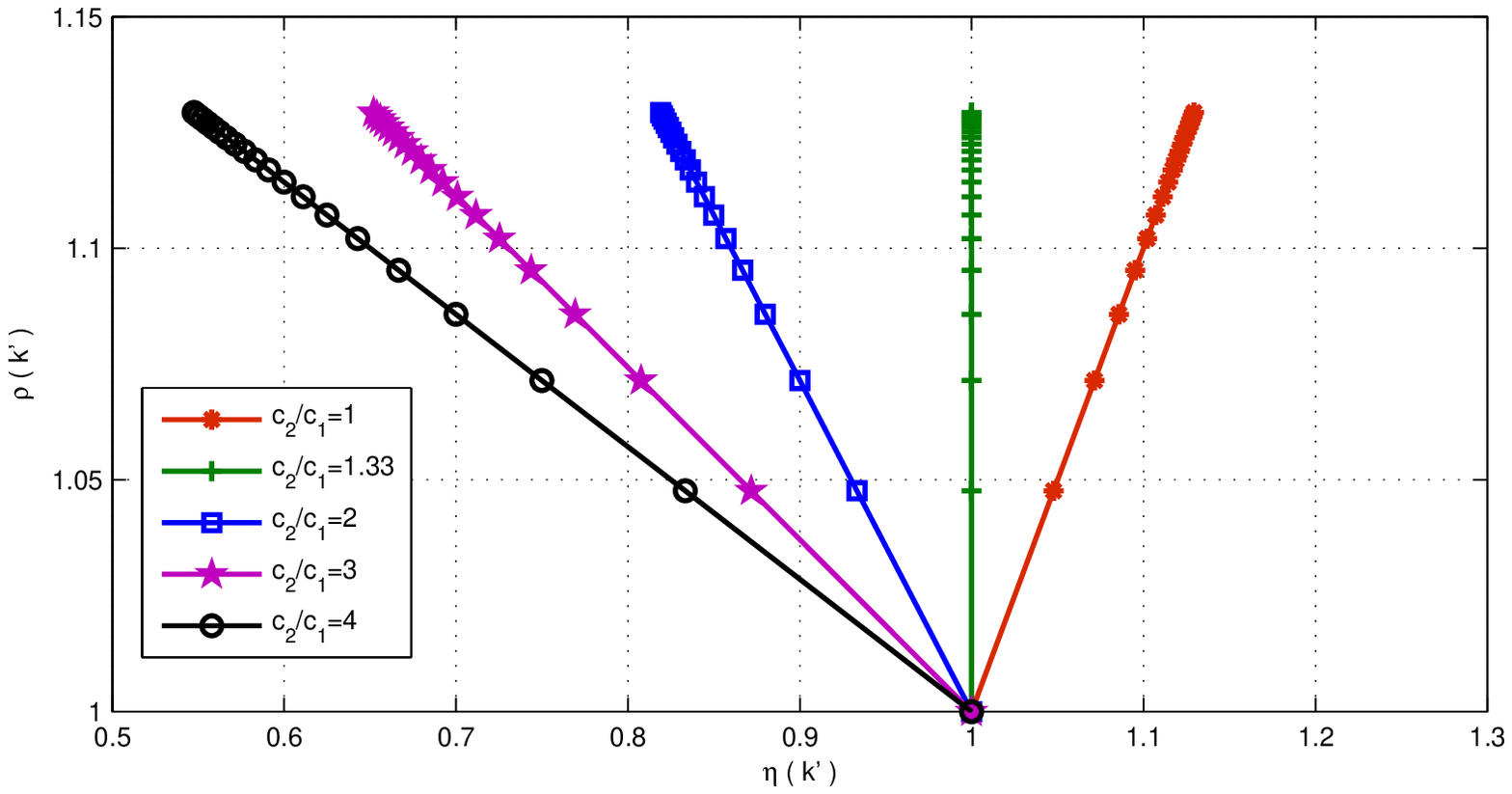,width=1\linewidth,clip=}
\caption{The tradeoff curves between the relative cost and repair bandwidth ratio for GMBR code.}
\label{fig:3}
\end{figure}

This section aims at providing some numerical results to get an insight regarding the proposed GMSR and GMBR codes and their advantages in terms of the corresponding storage capacity and/or repair bandwidth as compared to the MSR and MBR codes. In Fig.\ref{fig:2}, $\rho(k')$ versus $\eta(k')$ of the GMSR code for different integer values of $k'$ in the interval $[1,20]$ and for different relative cost ratios of $\frac{C_2}{C_1}$ is illustrated. Moreover, it is assumed $(n,k,d_1,d_2)=(15,5,8,6)$, which corresponds to scenario A, since $d_1\geq k$. Noting the condition (\ref{equ10}), in this example, if $\frac{C_2}{C_1}\geq\frac{d_1}{d_1-k+1}=2$, the download cost of GMSR is lower than that of MSR ($\eta(k')\leq 1$). This is in accordance to what is inferred from Fig.\ref{fig:2}. Moreover, Fig.\ref{fig:2} depicts the amount of increment in repair bandwidth for a given download cost ratio. Similarly, Fig.\ref{fig:3} provides the same result for GMBR with the same parameters, i.e., $(n,k,d_1,d_2)=(15,5,8,6)$. Again, referring to equation (\ref{equ14}), $\eta(k')\geq 1$ for $\frac{C_2}{C_1}\geq \frac{2d_1}{2d_1-k+1}=1.33$ which is in accordance to the result of Fig.\ref{fig:3}.

\begin{figure}
\centering
\epsfig{file=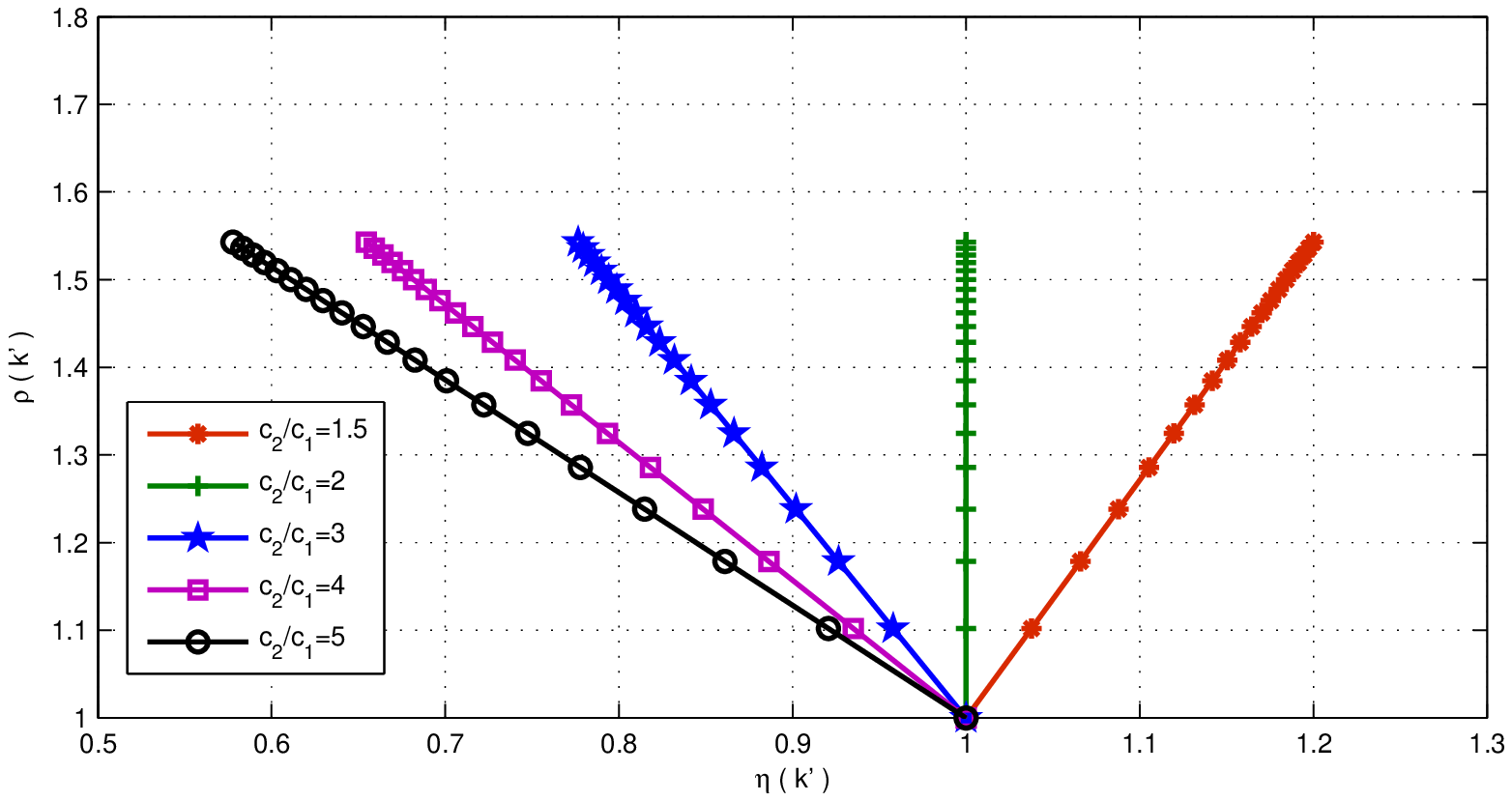,width=1\linewidth,clip=}
\caption{The tradeoff curves between the relative cost and bandwidth ratio for GMBR code.}
\label{fig:7}
\end{figure}

Fig.\ref{fig:7} depicts the $\rho(k')$ versus $\eta(k')$ for GMBR when $(n,k,d_1,d_2)=(15,5,4,10)$. Noting $d_1<k$, this case belongs to scenario B. Referring to (\ref{equ29}), if $\frac{C_2}{C_1}>\frac{2kd-k^2+k-d_1^2-d_1}{d_2(d_1+1)}=2$, the downlod cost of GMBR is lower than that of MBR ($\eta(k')\leq 1$). Fig.\ref{fig:7} confirms this threshold for $\frac{C_2}{C_1}$. Moreover, it shows how download cost ratio
affects the repair bandwidth ratio \big($\rho(k')$\big).

\begin{figure}
\centering
\epsfig{file=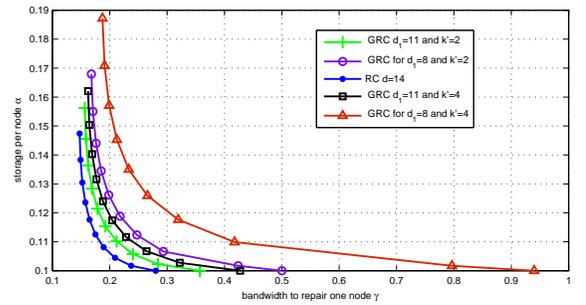,width=1\linewidth,clip=}
\caption{The tradeoff curves between the storage per node and repair bandwidth.}
\label{fig:5}
\end{figure}

Also, the tradeoff curves between the storage capacity per node and repair bandwidth for RC and GRC codes for two different values of $k'=2,4$ are shown in Fig.\ref{fig:5}. This shows the storage capacity-repair bandwidth tradeoff curve of RC code outperforms that of GRC (the dotted curve), while as is noted before, GRC may result in lower download cost as compared to that of RC code.

\begin{figure}
\centering
\epsfig{file=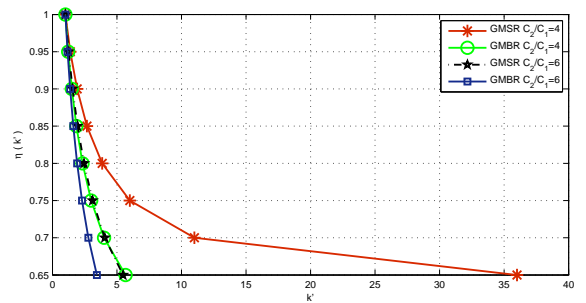,width=1\linewidth,clip=}
\caption{The effect of k' on the relative cost.}
\label{fig:6}
\end{figure}

Finally, Fig.\ref{fig:6} is provided to show the impact of different values of $k'$ on $\eta$ and for different values of $\frac{C_2}{C_1}$. Fig.\ref{fig:6} confirms that under certain conditions as is mentioned in the preceding sections, $\eta(k')$ is a decreasing function with respect to $k'$.

\section{Conclusion}\label{sec:conclusion}
This paper aims at addressing the cost bandwidth tradeoff in distributed storage systems when the download cost of storage nodes are not the same. Specifically, we concentrate to case that there are two sets of nodes, each having different download costs. Accordingly, using the corresponding \emph{Information Flow Graph}, a new variation of regenerating codes, called generalized regenerating codes, is proposed and is shown under some certain conditions outperform the current regenerating codes in terms of having lower download cost, while having a marginal increase in the repair bandwidth.

\bibliographystyle{IEEEtran}
\bibliography{refs}

\section{Appendix}\label{sec:appendix}
To derive the optimal tradeoff between $\alpha$ and $\beta_2$, one can fix $\beta_2$ and $d_1,d_2,k'$ (to some integer values) and then find the minimum value of $\alpha$ such that~(\ref{equ4}) and (\ref{equ16}) are satisfied. To this end, we define $\alpha_{min}$ as follows,
\begin{eqnarray}\label{equ30}
&\alpha_\textrm{min}&\!\!\!\!\!\!\!\!\!\!(d_1,d_2,k',\beta_2)\triangleq \textrm{min}~\alpha \nonumber\\
&~~~~~~\textrm{subject~to}&: C~\geq~M~,
\end{eqnarray}
where depending to on the condition that $d_1\geq k$ or $d_1<k$ we have,
\begin{eqnarray}\label{equ31}
C&\triangleq&\sum_{i=0}^{k-1}\textrm{min}\{(d_1\beta_1+d_2\beta_2-i\beta_1),~\alpha\}~\textrm{for}~d_1\geq k\nonumber\\
C&\triangleq&\sum_{i=0}^{d_1}\textrm{min}\{(d_1k'+d_2-ik')\beta_2,~\alpha\}\nonumber\\
&+&\sum_{i=d_1+1}^{k-1}\textrm{min}\{(d_1+d_2-i)\beta_2,~\alpha\}~\textrm{for}~d_1<k
\end{eqnarray}

The result of $d_1\geq k$:\\
To prove~(\ref{equ5}), substituting $\beta_1=k'\beta_2$ in the corresponding $C$ (equation (\ref{equ31}) with $d_1\geq k$), it follows,
\begin{eqnarray}\label{equ30}
\lefteqn{C\triangleq\sum_{i=0}^{k-1}\textrm{min}\{(d_1\beta_1+d_2\beta_2-i\beta_1),~\alpha\}}\nonumber\\
&=&\sum_{i=0}^{k-1}\textrm{min}\{(d_1k'+d_2-ik')\beta_2,~\alpha\}~\geq~M~.
\end{eqnarray}
Thus, $C$ can be computed, assuming $\alpha$ belongs to one of the following intervals,
\begin{eqnarray}\label{equ32}
C(\alpha)=~~~~~~~~~~~~~~~~~~~~~~~~~~~~~~~~~~~~~~~~~~~~~~~~~~~~~~~~~~~~~~~\nonumber\\
\nonumber\\
\left\{\begin{array}{ll}
k\alpha\!\!~~~~~&\textrm{$\alpha$}\in\big[0,h(1)\beta_2\big]\\
&\nonumber\\
(k-1)\alpha+h(1)\beta_2\!\!~~~~~&\textrm{$\alpha$}\in\big(h(1)\beta_2,h(2)\beta_2\big]\\
&\nonumber\\\vdots
&\nonumber\\
(k-j)\alpha+\sum_{i=1}^{j}h(i)\beta_2\!\!~~~~~&\textrm{$\alpha$}\in\big(h(j)\beta_2,h(j+1)\beta_2\big]\\
&\nonumber\\\vdots
&\nonumber\\
\alpha+\sum_{i=1}^{k-1}h(i)\beta_2\!\!~~~~~&\textrm{$\alpha$}\in\big(h(k-1)\beta_2,h(k)\beta_2\big]~,\\
\end{array} \right.
\end{eqnarray}
where
\begin{eqnarray}\label{equ33}
h(i)\triangleq d_1k'+d_2-(k-i)k'
\end{eqnarray}
As a result, noting $C\geq M$, it follows,
\begin{eqnarray}\label{equ34}
\alpha_{\textrm{min}}=~~~~~~~~~~~~~~~~~~~~~~~~~~~~~~~~~~~~~~~~~~~~~~~~~~~~~~~~~~~~~~~\nonumber\\
&\nonumber\\
\left\{\begin{array}{ll}
\frac{M}{k}\!\!&~\textrm{$M$}\in\big[0,kh(1)\beta_2\big]\\
&\nonumber\\
\frac{M-\big(\sum_{i=1}^{j}h(i)\big)\beta_2}{(k-j)}\!\!&~\textrm{$M$}\in\big((k-j)h(j)\beta_2+\sum_{i=1}^{j}h(i)\beta_2,\nonumber\\
&~~(k-j)h(j+1)\beta_2+\sum_{i=1}^{j}h(i)\beta_2\big]~,\\
&~~~~~~~~~j=0,1,...,k-1
\end{array} \right.
\end{eqnarray}
or equivalently,
\begin{eqnarray}\label{equ35}
\alpha_{\textrm{min}}=~~~~~~~~~~~~~~~~~~~~~~~~~~~~~~~~~~~~~~~~~~~~~~~~~~~~~~~~~~~~~~~~~\nonumber\\
\nonumber\\
\left\{\begin{array}{ll}
\frac{M}{k}\!\!&\textrm{$\beta_2$}\in\big[\frac{M}{kh(1)},\infty\big)\\
&\nonumber\\
\frac{M-\big(\sum_{i=1}^{j}h(i)\big)\beta_2}{(k-j)}\!\!&~\textrm{$\beta_2$}\in\big[\frac{M}{(k-j)h(j+1)+\sum_{i=1}^{j}h(i)},\\
&\nonumber\\
&~\frac{M}{(k-j)h(j)+\sum_{i=1}^{j}h(i)}\big),~j=0,1,...,k-1\\
\end{array} \right.
\end{eqnarray}
As a result, noting (\ref{equ33}), it follows,
\begin{eqnarray}\label{equ36}
\alpha_{\textrm{min}}=~~~~~~~~~~~~~~~~~~~~~~~~~~~~~~~~~~~~~~~~~~~~~~~~~~~~~~~~~~~~~~~~~~~~\nonumber\\
&\nonumber\\
\left\{\begin{array}{ll}
\frac{M}{k}\!\!&\textrm{$\beta_2$}\in\big[f(0),\infty\big)\\
&\nonumber\\
\frac{2M-g(i)\beta_2}{2(k-i)}\!\!&\textrm{$\beta_2$}\in\big[f(i),f(i-1)\big)~,~~~~~~i=0,1,...,k-1,\\
\end{array} \right.
\end{eqnarray}
where
\begin{eqnarray}\label{equ37}
f(i)\triangleq \frac{2M}{2kh(0)+(i+1)(2k-i)k'}
\end{eqnarray}
\begin{eqnarray}\label{equ38}
g(i)\triangleq i(2d_1k'+2d_2-2kk'+(i+1)k')
\end{eqnarray}
\begin{eqnarray}\label{equ39}
\beta_{\textrm{min}}=f(k-1)
\end{eqnarray}

The result of $d_1< k$:\\

In this case, substituting $\beta_1=k'\beta_2$ in $C$ (Equation (\ref{equ31}) when $d_1<k$) and following the same approach as the case of $d_1\geq k$, it follows,
\begin{eqnarray}\label{equ42}
C(\alpha)=~~~~~~~~~~~~~~~~~~~~~~~~~~~~~~~~~~~~~~~~~~~~~~~~~~~~~~~~~~~~~~~~~~~\nonumber\\
\nonumber\\
\left\{\begin{array}{ll}
k\alpha\!\!\!\!\!\!&\textrm{$\alpha$}\in[0,h(1,0)\beta_2]\\
(k-1)\alpha+h(1,0)\beta_2\!\!\!\!\!\!&\textrm{$\alpha$}\in(h(1,0)\beta_2,h(2,0)\beta_2]\\
\vdots\\
(k-j)\alpha+\sum_{i=1}^{j}{h(i,0)\beta_2}\!\!\!\!\!\!~&\textrm{$\alpha$}\in(h(j,0)\beta_2,h(j+1,0)\beta_2]\\
\vdots\\
d_1\alpha+\sum_{i=1}^{k-d_1}{h(i,0)\beta_2}\!\!\!\!\!\!&\textrm{$\alpha$}\in(h(k-d_1,0)\beta_2,\\
&~~~~~~~h(k-d_1,1)\beta_2]\\
A\!\!\!\!&\textrm{$\alpha$}\in(h(k-d_1,1)\beta_2,\\
&~~~~~~~h(k-d_1,2)\beta_2]\\
\vdots\\
B&\textrm{$\alpha$}\in(h(k-d_1,t)\beta_2,\\
&~~~~~~~h(k-d_1,t+1)\beta_2]\\
\vdots\\
D\!\!\!\!\!\!\!\!\!\!\!\!\!\!\!\!\!\!\!\!\!\!\!\!\!\!\!\!\!\!\!\!\!\!\!\!\!\!\!\!\!\!\!&\textrm{$\alpha$}\in(h(k-d_1,d_1-1)\beta_2,\nonumber\\
&~~~~~~~h(k-d_1,d_1)\beta_2]\\
\end{array} \right.
\end{eqnarray}
where
\begin{eqnarray}\label{equ43}
h(x,y)&\triangleq& d_1+d_2-k+x+yk'\nonumber\\
A&=&(d_1-1)\alpha+\sum_{i=1}^{k-d_1}{h(i,0)\beta_2}+h(k-d_1,1)\beta_2\nonumber\\
B&=&(d_1-t)\alpha+\sum_{i=1}^{k-d_1}{h(i,0)\beta_2}+\sum_{i=1}^{t}{h(k-d_1,i)\beta_2}\nonumber\\
D&=&\alpha+\sum_{i=1}^{k-d_1}{h(i,0)\beta_2}+\sum_{i=1}^{d_1-1}{h(k-d_1,i)\beta_2}\nonumber\\
C(\alpha_{\textrm{min}})&=&M~.
\end{eqnarray}
Thus, $\alpha_{\textrm{min}}=C^{-1}(M)$ can be computed as,
\begin{eqnarray}\label{equ44}
\alpha_{\textrm{min}}(d_1,d_2,k',\beta_2)=~~~~~~~~~~~~~~~~~~~~~~~~~~~~~~~~~~~~~~~~~\nonumber\\
\nonumber\\
\left\{\begin{array}{ll}
\frac{M}{k}\!\!&\textrm{$\beta_2$}\in\big[f_1(0),\infty\big)\\
&\nonumber\\
\frac{2M-g_1(i)\beta_2}{2(k-i)}\!\!&\textrm{$\beta_2$}\in\big[f_1(i),f_1(i-1)\big)\\
&\nonumber\\
\frac{2M-(g_1(k-d_1-1)+g_2(i))\beta_2}{2(d_1-i)}\!\!&\textrm{$\beta_2$}\in\big[f_2(i),f_2(i-1)\big)~,\\
\end{array} \right.
\end{eqnarray}
where
\begin{eqnarray}\label{equ45}
f_1(i)\!\!&\triangleq&\!\! \frac{2M}{2k(d-k)+(i+1)(2k-i))}\nonumber\\
f_2(i)\!\!&\triangleq&\!\! \frac{2M}{(2kd-k^2-d_1^2-d_1+k+2d_1k')+ik'(2d_1-i-1)}\nonumber\\
g_1(i)\!\!&\triangleq&\!\! i(2d-2k+i+1)\nonumber\\
g_2(i)\!\!&\triangleq&\!\! (i+1)(2d_2+ik')~.\!\!\!\!\!\!\!\!\!\!\!\!\!\!\!\!
\end{eqnarray}
Finally, $\beta_{2{\textrm{min}}}$ can be computed as,
\begin{equation}\label{equ46}
\!\!\!\!\!\!\!\!\beta_{2_{min}}=f_2(d_1-1)
\end{equation}
\end{document}